\documentstyle[12pt,epsf]{article}
\newcommand{\lsim}{\,{\buildrel < \over {_\sim}}\,}

\begin{document}

\begin{titlepage}
\begin{flushright}
JYFL-7/99\\
LPT Orsay 99-48\\
hep-ph/9906484
\end{flushright}
\vspace{3cm}

\begin{centering}

{\Large \bf Nuclear Parton Distributions - a DGLAP Analysis\footnote{
Talk given by K.J. Eskola in {\em Quark Matter '99}, 12 May, 1999, 
Torino, Italy. For the transparencies, see 
http://www.qm99.to.infn.it/program/qmprogram.html}}

\vspace{0.5cm}
{\large K.J. Eskola, V.J. Kolhinen, P.V. Ruuskanen\\}
\vspace{0.3cm}
{\em Department of Physics, University of Jyv\"askyl\"a,\\
P.O.Box 35, FIN-40351 Jyv\"askyl\"a, Finland\\
e-mail: kari.eskola,vesa.kolhinen,vesa.ruuskanen@phys.jyu.fi\\ 
}
\vspace{0.3cm}
{\large  C.A. Salgado}
\vspace{0.3cm}

{\em Laboratoire de Physique Th\'eorique, Universit\'e de Paris XI,\\
 Batiment 211, F-91405 Orsay Cedex, France\\
e-mail: salgado@qcd.th.u-psud.fr\\}

\vspace{1cm}
\end{centering}

\begin{abstract}
{\normalsize Nuclear parton distributions $f_A(x,Q^2)$ are studied within a
framework of the DGLAP evolution. Measurements of $F_2^A/F_2^D$ in deep
inelastic $lA$ collisions, and Drell--Yan dilepton cross sections
measured in $pA$ collisions are used as constraints. Also conservation
of momentum and baryon number is required. It is shown that the
calculated $Q^2$ evolution of $F_2^{\rm Sn}/F_2^{\rm C}$ agrees very
well with the recent NMC data, and that the ratios $R_f^A=f_A/f$ are
only moderately sensitive to the choice of a specific modern set of
free parton distributions. For general use, we offer a numerical
parametrization of $R_f^A(x,Q^2)$ for all parton flavours $f$ in
$A>2$, and at $10^{-6}\le x \le 1$ and $2.25\, {\rm GeV}^2\le Q^2\le
10^4\,{\rm GeV}^2$.
}
\end{abstract}
\end{titlepage}

\section{The Framework}

The motivation to study nuclear parton distributions are the hard
probes of strongly interacting matter \cite{HPC}. Due to a large
momentum (or mass) scale $Q\gg\Lambda_{\rm QCD}$ involved, these
processes take place during the very first fractions of fm/$c$ in the
high energy nuclear collisions, acting as probes of the forming 
quark gluon plasma. Due to the large scale, hard
processes are computable within QCD perturbation theory. To a first
approximation the cross sections of hard processes in nuclear
collisions can be factorized as in hadronic collisions, into parton
densities and a hard parton-parton cross section. The parton
distributions in bound and free protons, however, are different:
$f_A(x,Q^2)\ne f(x,Q^2)$. Typically in hard collisions 
$x\sim Q/\sqrt s$, so quite different regions in $x$ become relevant 
when moving from the present SPS-energies $\sqrt s/A \sim
20$ GeV up to the future LHC energies $\sqrt s/A \sim 5.5$ TeV.  Thus
there is a need for analyses of nuclear parton distributions which
consistently cover sufficiently wide ranges in $x$ and $Q^2$. In this
talk, I will present the main results from such an analysis
\cite{EKR,EKS}.

In deeply inelastic $lA$ scatterings (DIS), ratios of measured
differential cross sections, $\frac{1}{A}\frac{d\sigma^{lA}}{dxdQ^2}/
\frac{1}{2}\frac{d\sigma^{lD}}{dxdQ^2}$, reflect the corresponding
ratios of the nuclear structure function $F_2^A$ and that of deuterium
$F_2^D$. The ratio $F_2^A/F_2^D$ is observed to deviate clearly from
unity \cite{ARNEODO94}. Since $F_2(x,Q^2) = \sum_q e_q^2 [xq(x,Q^2) +
x\bar q(x,Q^2)]$, parton distributions in bound protons obviously
differ from those in the free proton. Often the nuclear modifications are
referred to as shadowing $(x \lsim 0.1)$, anti-shadowing $(0.1 \lsim x
\lsim 0.3)$, EMC effect $(0.3 \lsim x\lsim0.7)$ and Fermi motion
($x\rightarrow1$ and beyond). The dependence on the Bjorken $x$ has
been observed already in the 80's \cite{ARNEODO94} but the weaker
$Q^2$ dependence was detected only fairly recently by the NMC
\cite{NMC96}.

In Refs. \cite{EKR,EKS} our goal has {\em not} been to study the
actual origin of the modifications but, rather - because perturbative
QCD (pQCD) does not predict the absolute parton distributions - to use
the observed effects as {\em input} for an analysis in the
perturbative region. The basic idea in our study is the same as in the
global analyses of parton distributions of the free proton (like in
Ref. \cite{CTEQ}): we determine the nuclear parton densities at a wide
range of $x$ and $Q\ge Q_0\gg\Lambda_{\rm QCD}$ through their
perturbative QCD (DGLAP \cite{DGLAP}) evolution by using available
experimental data and conservation of momentum and baryon number as
constraints.

Information of the nuclear parton distributions can be obtained from
$lA$ DIS and Drell-Yan (DY) measurements in $pA$ collisions.  In these
measurements, the accessible values of $x$ and $Q^2$ are
strongly correlated, as illustrated in Fig. 1. To perform the DGLAP
evolution of the parton densities, however, the initial distributions
are needed along a fixed scale $Q_0^2$.  Therefore, we determine the
initial nuclear parton distributions at $Q_0^2$ {\it iteratively},
through the DGLAP evolution, by using the available data at scales
$Q^2\ge Q^2_0$ as constraints. Note that now the problem is more
complicated than in the free proton case because of the additional
variable $A$.

\begin{figure}[htb]
\vskip -1.0cm 
\hfill \epsfxsize=9cm\epsfbox{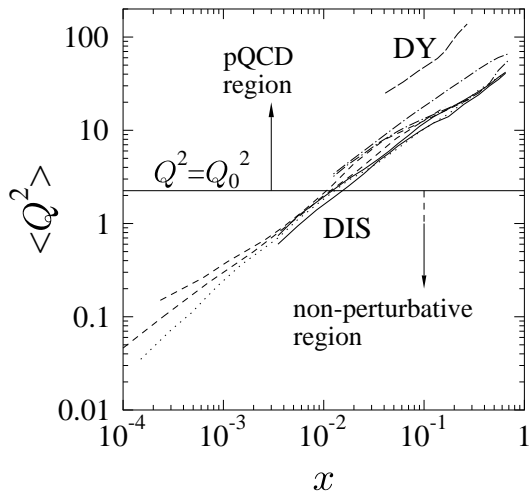}
\vskip -7cm 
\caption{ The correlation of $x$ and  $\langle Q^2\rangle$ {\protect\\} 
in the measurements of DIS {\protect\cite{NMC,E665}} in $lA$  {\protect\\} 
and DY ($x=x_2$) {\protect\cite{E772}}  in $pA$. Our choice {\protect\\}
for the  initial scale $Q_0^2$ is also indicated.
 \label{fig1}}
\vskip 2.cm 
\end{figure}

We take the parton distributions of the free proton as accurately
known.  We choose $Q_0^2=2.25$ GeV$^2$ which is the $c$-quark mass
threshold in the set GRVLO \cite{GRVLO} we are using as the basis. We
first parametrize the ratio $R_{F_2}^A(x,Q_0^2)$ for (isoscalar) $A$
and $x$. The potential but small nuclear effects in deuterium, and the
small tails at $x>1$ are neglected here. At the initial scale $Q_0^2$
(but only at $Q_0^2$), we assume that the nuclear sea quarks and
antiquarks are modified approximately with the same profile, and
similarly for the valence quarks. The ratio $F_2^A/F_2^D$ can then be
simply written as a linear combination of a sea quark ratio
$R_S^A=S_A/S$ and a valence quark ratio $R_V^A=V_A/V$, where $S\,(V)$
are the total sea (valence) distributions. The ratios $R_S^A$ and
$R_V^A$ at $Q_0^2$ in turn are constrained by the DIS data
\cite{NMC,SLACre,E665} and DY data in $pA$ collisions \cite{E772} (see
Fig. 1) at higher scales. $R_V^A$ is also further constrained by
baryon number conservation. Momentum conservation gives an overall
constraint for the gluon distributions at $Q_0^2$. In lack of any
direct constraints for the nuclear gluons from the data, we assume
that initially $g_A/g\approx R_S^A$ at very small values of $x$. The
value of $x$ where $R_G^A(x,Q_0^2)=1$ is estimated on the basis of
Ref. \cite{PIRNER}. For more details, please see Ref. \cite{EKR}. Once
the initial distributions for all parton flavours have been determined
like this, the DGLAP evolution to larger scales can be performed, and
comparison with the data can be made (see Fig.1). The initial ratios
$R_S^A(x)$, $R_V^A(x)$, $R_G^A(x)$ at $Q^2_0$ are then iterated
until a ``best'' initial condition is found.

\begin{figure}[hbt]
\centerline{{\small \hskip 1cm \bf (a)\hskip7.5cm (b)}\hskip 2cm}
\vskip -1cm
\centerline{\hskip -1.9cm
\epsfxsize=10cm\epsfbox{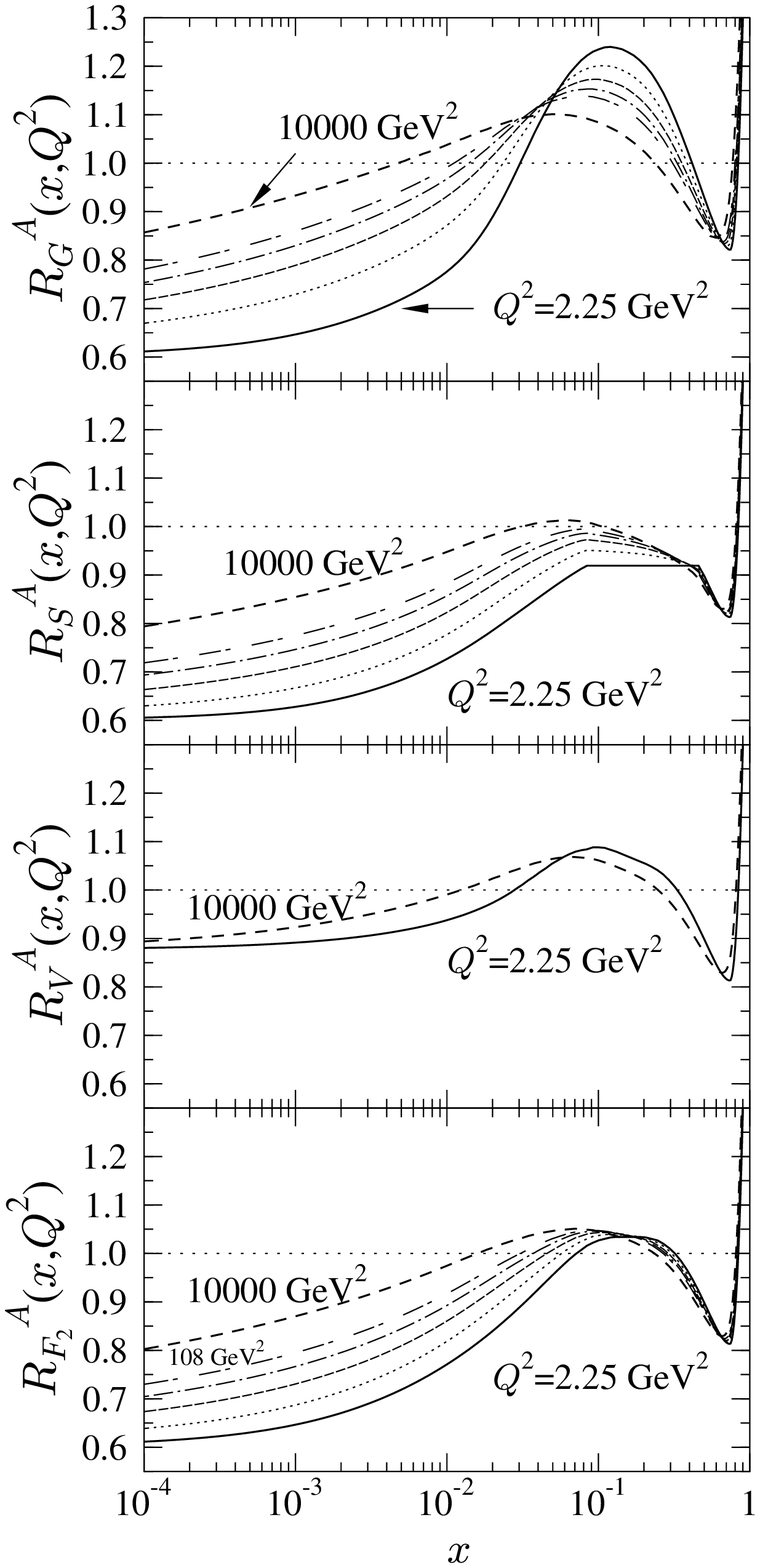} 
\hskip -1.0cm
\epsfxsize=10cm\epsfbox{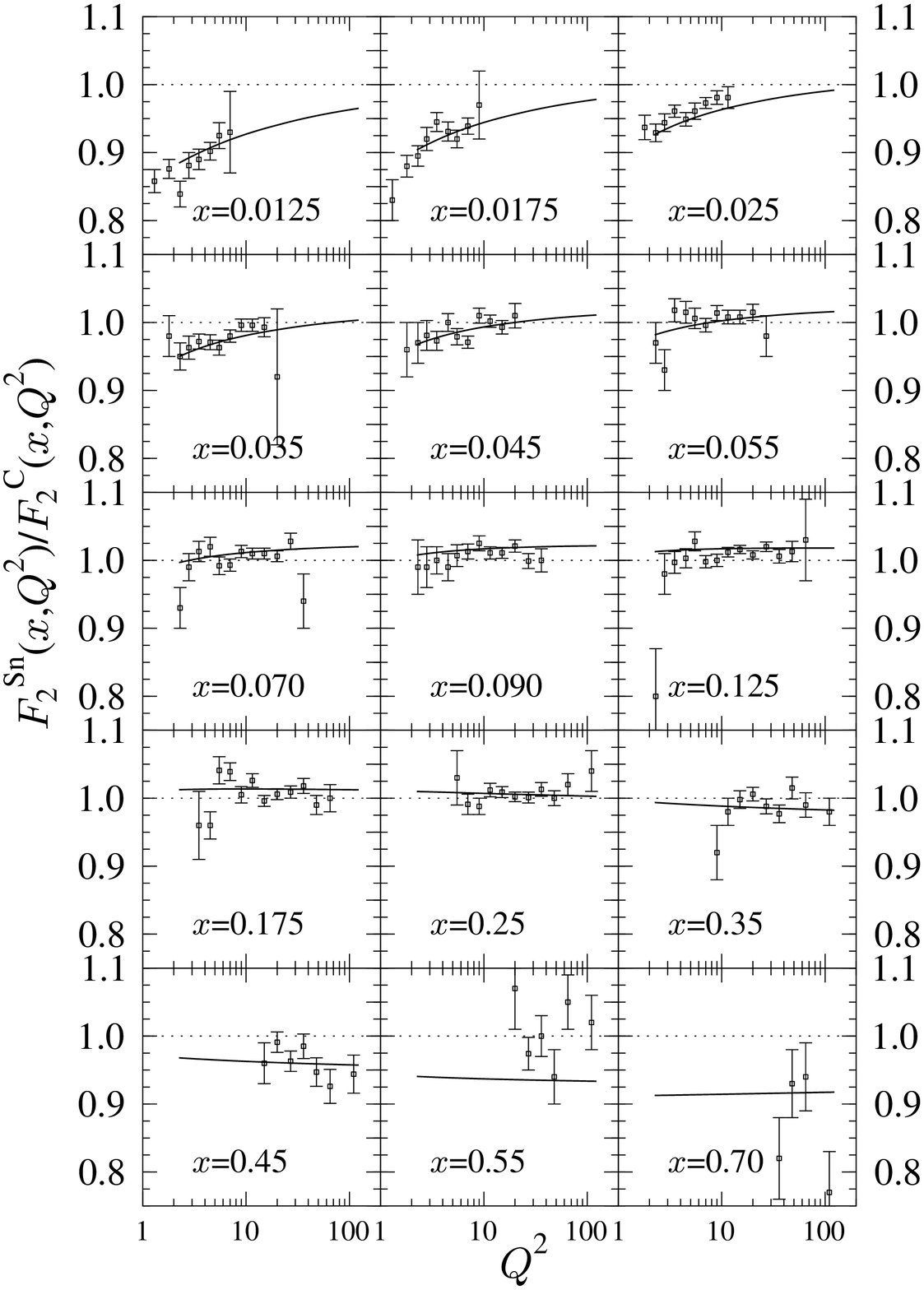}
}
\vskip -.5cm
\caption{\small (a) The scale evolution of the ratios $xg_A/xg$, $xS_A/xS$, $xV_A/xV$ 
and $F_2^A/F_2^D$ for an isoscalar nucleus $A=208$.
(b) The calculated $Q^2$ dependence of $F_2^{\rm Sn}/F_2^{\rm C}$ 
compared with the NMC data {\protect\cite{NMC96}}.  
\label{fig:fig2} }
\end{figure}

Fig. 2a shows the scale evolution of the nuclear effects in parton
distributions for an isoscalar nucleus $A$=208. The ratios $g_A/g$,
$S_A/S$, $V_A/V$ and $F_2^A/F_2^D$ are shown as functions of $x$ at
fixed values of $Q^2=$ 2.25~GeV$^2$ (solid lines), 5.39~~GeV$^2$
(dotted), 14.7~~GeV$^2$ (dashed), 39.9~~GeV$^2$ (dotted-dashed),
108~GeV$^2$ (double-dashed), equidistant in $\log Q^2$, and
10000~GeV$^2$ (dashed). For $R_V^A$ only the first and last ones are
shown.

In Fig. 2b we plot the calculated lowest order, leading twist pQCD
evolution for the ratio $F_2^{\rm Sn}/F_2^{\rm C}$ at different fixed
values of $x$, and compare the results directly with the corresponding
data of NMC \cite{NMC96}. The agreement is very good.  Note that in
order to reduce the potential gluon fusion \cite{GLRMQ} effects in the
evolution as much as possible, we have chosen the initial scale above
1 GeV but below the $m_c$-threshold in order to make the treatment of
the initial state as simple as possible. The $\log Q^2$ slopes of
$F_2$, and also of the ratio $F_2^A/F_2^D$, can be used to constrain
the gluon distributions \cite{PIRNER}. Unfortunately, however, the
values of $x$ of the NMC data \cite{NMC96} are not quite small enough
to get a firm handle on the initial nuclear shadowing of gluons.

\section{The EKS98-parametrization}

We have also repeated the analysis by using the CTEQ4L parton
distributions \cite{CTEQ} as the basis \cite{EKS}. Note that in CTEQ4L
the sea is more flavour-asymmetric than in GRVLO, and that there are
less gluons in CTEQ4L at very small values of $x$.  The nuclear ratios
$R_f^A\equiv f_A/f$ for each flavour of partons were, however, found
to deviate at most a few per cents relative to those computed with
GRVLO. We therefore conclude that to a good first approximation
$f_A(x,Q^2)_{\rm set}=R_f^A(x,Q^2)f(x,Q^2)_{\rm set}$, where ``set''
refers to any modern lowest order set of parton distributions for the
free proton, and where $R_f^A$ does {\em not} depend on the set.

For practical applications of computing hard cross sections in high
energy nuclear collisions, we have also prepared a parametrization of
the nuclear ratios $R_f^A(x,Q^2)$ for each parton flavour $f$ in any
nucleus $A>2$, at $10^{-6}\le x \le 1$ and $2.25\, {\rm GeV}^2\le
Q^2\le 10^4\,{\rm GeV}^2$.  The parametrization is intended for
general use, and it is available from us via email, or from
http://fpaxp1.usc.es/phenom/ or from http://www.urhic.phys.jyu.fi/.

Finally, we note that our analysis can be improved in obvious ways:
more quantitative error analysis must be implemented, next-to-leading
order DGLAP evolution must be done, the initial gluon distributions
and gluon fusion corrections \cite{GLRMQ} must be studied in more
detail.
\vskip 1.cm

\noindent{\bf Acknowledgments.} 
We thank the Academy of Finland and C.A.S. thanks Fundaci\'on Caixa
Galicia from Spain for financial support.

\end{document}